\documentclass{emulateapj}

\usepackage{graphicx}
\usepackage[space]{grffile}
\usepackage{latexsym}
\usepackage{amsfonts,amsmath,amssymb}
\usepackage{url}
\usepackage[utf8]{inputenc}
\usepackage{fancyref}
\usepackage{hyperref}
\hypersetup{colorlinks=false,pdfborder={0 0 0},}
\usepackage{textcomp}
\usepackage{longtable}
\usepackage{multirow,booktabs}

\usepackage{natbib}

\usepackage{supertabular}

\usepackage[caption=false]{subfig}

\usepackage{calrsfs}
\DeclareMathAlphabet{\pazocal}{OMS}{zplm}{m}{n}
\newcommand{\La}{\mathcal{L}}

\def\mpsp{m\,s$^{-1}$}

\def\ha{H$\alpha$~}

\def\shkp{$S_{HK}$}

\def\kep{\emph{Kepler}~}
\def\kep{\emph{Kepler}}

\shorttitle{An Ultra-Short Period for YZ Ceti c?}
\shortauthors{Robertson}

\begin{document}


\title{Aliasing in the Radial Velocities of YZ Ceti: An Ultra-Short Period for YZ Ceti c?}


\author{Paul Robertson}

\affil{Department of Physics \& Astronomy, The University of California, Irvine, 4129 Frederick Reines Hall, Irvine, CA, 92697, USA}
\email{paul.robertson@uci.edu}


\begin{abstract}

Mid-late M stars are opportunistic targets for the study of low-mass exoplanets in transit because of the high planet-to-star radius ratios of their planets.  Recent studies of such stars have shown that, like their early-M counterparts, they often host multi-resonant networks of small planets.  Here, we reanalyze radial velocity measurements of YZ Ceti, an active M4 dwarf for which the HARPS exoplanet survey recently discovered three exoplanets on short-period (P = 4.66, 3.06, 1.97 days) orbits.  Our analysis finds that the orbital periods of the inner two planets cannot be uniquely determined using the published HARPS velocities.  In particular, it appears likely that the 3.06-day period of YZ Ceti c is an alias, and that its true period is 0.75 days.  If so, the revised minimum mass of this planet is less than 0.6 Earth masses, and its geometric transit probability increases to 10\%.  We encourage additional observations to determine the true periods of YZ Ceti b and c, and suggest a search for transits at the 0.75-day period in TESS lightcurves.

\end{abstract}



\section{Introduction}

Mid-late M stars are increasingly common targets of exoplanet surveys.  \kep \citep{borucki10} included relatively few such stars in its target list, but its extended mission K2 has revealed several systems of small planets orbiting very low-mass stars \citep[e.g.][]{mann16,hirano16}.  TESS \citep{ricker15} has started science operations, and will add many more systems to the catalog of exoplanets around late-type stars.  At the same time, a collection of near-infrared Doppler spectrographs is going into operation, beginning with CARMENES \citep{quirrenbach16} and HPF \citep{mahadevan14}, which will enable ground-based follow-up to determine the masses of planets transiting these cool, faint stars.

Already, several of the most high-profile recent exoplanet discoveries have been around mid-late M stars.  TRAPPIST-1, a nearby M8 dwarf, was shown to host a multi-resonant network of seven low-mass exoplanets \citep{gillon17}, three of which lie within the liquid-water habitable zone \citep[HZ;][]{kopparapu13}.  Radial velocity (RV) surveys have also discovered low-mass exoplanets around nearby mid-late M stars.  \citet{anglada16} found evidence for a terrestrial-mass planet in the HZ of Proxima Centauri.  More recently, \citet{astudillo-defru17} announced the discovery of three Earth-mass exoplanets orbiting the M4.5 dwarf YZ Ceti based on observations from the HARPS spectrograph.  At candidate periods of 1.97, 3.06, and 4.66 days, the YZ Ceti system potentially represents another compact multi-harmonic system like TRAPPIST-1.  TESS will observe YZ Ceti in late 2018, and all of the reported planets have relatively high ($P \sim 5$\%) geometric transit probabilities.

Ground-based exoplanet surveys are plagued by difficulties associated with temporal sampling.  Uneven time sampling caused by shared telescope resources, seasonal target observability, weather, and the day/night cycle limit sensitivity in certain regions of frequency space, and can create ambiguities in others.  Aliasing occurs when a continuous signal is observed at a cadence such that the observations cannot distinguish between the true signal frequency and a combination of the signal and observing frequencies.  RV surveys are commonly hampered by the ``1-day alias", and periodograms of RV data will often show peaks at the frequency of a planet ($f_p$) and its alias at $f_a = f_p \pm 1$ day$^{-1}$.  This effect was demonstrated most powerfully by \citet{dawson10}, who revised the period of 55 Cnc e from 2.8 days \citep{mcarthur04} to its true value of 0.75 days, where it was later found to transit \citep{winn11}.

In this Letter, we argue that the periods of two of the three planets orbiting YZ Ceti are not well determined due to aliasing in the HARPS RV time series.  For planet b, which has a period of either 1.97 or 2.02 days, the difference is primarily important for the efficiency of identifying potential transits.  On the other hand, the period of planet c may be 0.75 days rather than 3.06 days, which significantly alters its derived physical properties and geometric transit probability.  Given that the available RVs are unable to clearly distinguish between these candidate periods, it will be especially important to examine all potential transit windows in TESS lightcurves of YZ Ceti.

\section{Data and Analysis}

\subsection{Data}

In this Letter, we have analyzed the HARPS observations of YZ Ceti as presented in Table B.4 of \citet{astudillo-defru17}.  The quantities derived from time-series spectroscopic observations of YZ Ceti include RVs, as well as spectral properties sensitive to stellar magnetic activity.  The activity tracers include the full width at half maximum (FWHM) of the cross-correlation function, the line bisector slopes (BIS), and strengths of the calcium H\&K (\shkp) and \ha absorption lines.

\subsection{Stellar Activity}

\citet{astudillo-defru17} analyzed photometry of YZ Ceti from the All-Sky Automated Survey \citep[ASAS;][]{pojmanski97} and the HARPS FWHM values, finding evidence of the stellar rotation period at $P_{rot} \sim 83$ days.  They find no evidence for this period in the RV data or the absorption-line activity indicators, an assessment with which we agree.  However, we note that the \ha time series (and \shkp, at lower S/N) does include several interesting periodicities, including at periods near 500 and 53 days, and possibly also a long-period trend.  These periods are difficult to interpret in light of the candidate rotation period at 83 days.  The 500-day period is shorter than typical stellar magnetic cycles, but could be a ``sub-cycle" of the longer-term magnetic evolution indicated by the trend.  Similar behavior has been observed for the Sun \citep[e.g.][]{wauters16}.  It is possible that the rotation period is either 53 or 83 days, and the other period is a typical active region lifetime.

None of the periods identified in the stellar activity indicators appear at significant power in RV.  Depending on the model adopted for the planets, we sometimes observe residual power near the $\sim 25$-day harmonic of the 53-day period, but at levels far too low to be statistically significant.

Thus, it appears that the timescales for the primary stellar activity signals, as well as their dominant harmonics, lie far from the periods of the candidate exoplanets.  \citet{astudillo-defru17} included a Gaussian process (GP) correlated noise component \citep{rasmussen05} in their 3-planet model, but we see no evidence for correlated noise from astrophysical variability near the planets under consideration.  We find that the derived properties of the planets do not change significantly when including a GP noise model, and that we cannot meaningfully constrain the hyperparameters of the quasi-periodic GP kernel.  Thus, for our analysis, we have modeled the HARPS time series as a sum of three Keplerian functions.

\subsection{RV Period Search}

We sought to identify periodicity in the RV time series using three periodograms, each with different advantages.  We first used the traditional Lomb-Scargle periodogram (GLS), as fully generalized by \citet{zk09}.  We also considered the Bayes factor periodogram (BFP), provided in the \texttt{Agatha} software suite by \citet{feng17}.  The BFP computes the power spectrum by comparing the Bayesian information criterion (BIC) for a periodic signal to that of the noise model at each period, and includes options for correlated noise models and correlations with activity proxies.  Finally, we computed the compressed sensing periodogram (CSP) as described by \citet{hara17}.  Whereas the GLS and BFP evaluate one period at a time, and identify multiple signals iteratively by removing the strongest signal and recomputing, the CSP models the entire frequency parameter space simultaneously by fitting amplitudes to a large library of periodic signals (here, sinusoids).  By evaluating all periods simultaneously, the CSP excels at minimizing the impact of aliasing.

We note that the model selection routine in \texttt{Agatha} prefers a white noise model with no correlations to activity proxies for the RV series.  Thus, the GLS and BFP power spectra are largely similar.  However, the BFP offers the advantage of a more robust threshold for statistical significance.  Namely, as discussed in \citet{feng17}, peaks with power $\ln(BF) > 5$ are generally considered significant.

\begin{figure}
\centering
\includegraphics[width=\columnwidth]{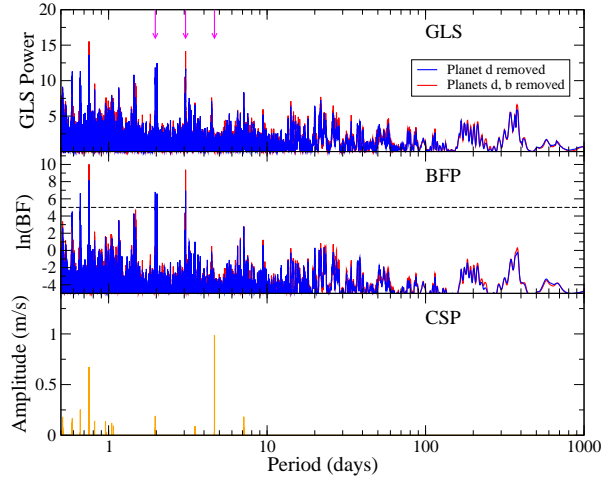}
\caption{\footnotesize Periodograms of the HARPS RVs of YZ Ceti.  For the GLS and BFP periodograms, we show the power spectra after successively modeling and subtracting planets d (\emph{blue}) and b (\emph{red}).  The pink arrows indicate the periods of the three planets as published by \citet{astudillo-defru17}.  All three periodograms prefer the 0.75-day period for planet c over its 1-day alias at 3.06 days.}
\label{fig:periodograms}
\end{figure}

All three periodograms identify the 4.66-day signal of YZ Ceti d as the strongest periodicity in the RV time series.  However, subsequent analysis of the power spectra reveals ambiguities for the periods of each of the other two proposed planets in the system.  In Figure \ref{fig:periodograms}, we show our periodograms.  For the GLS and BFP periodograms, we have successively modeled and subtracted the orbits of planet d and b\footnote{Ordinarily, the BFP automatically models and subtracts the strongest periodogram peak at each step.  For the purpose of determining the period of planet c, we have manually removed planets d and b even though the peaks associated with planet c are stronger.} in order to study the residual periodicities.  The three periodograms again agree on the second-strongest signal, this time at a period of 0.75 days.  This period is a 1-day alias of the 3.06-day period attributed to planet c by \citet{astudillo-defru17}.  The GLS and BFP power spectra also show significant power at the 3.06-day period, while the CSP converges on a single model that prefers the 0.75-day period.

Furthermore, the GLS and BFP periodograms indicate two possible periods for planet b, one at the published value of 1.97 days, and one at a slightly longer period of 2.02 days.  Again, the 2.02-day period is a 1-day alias of the 1.97-day period.  While the physical properties of planet b are minimally dependent on such a small difference in period, it will be important to identify the correct period for future transit searches.

The CSP does not recover the 2-day planet at any significant amplitude.  In general, we find that the detections of planets b and c are marginal, and strongly dependent on the RVs from the first season of high-cadence observations in 2013.  The stellar activity indicators suggest YZ Ceti is relatively quiet during this season, and we do not observe periodicity near the planet periods in the activity tracers when isolating the 2013 observations.  Thus, there is no particular reason to exclude or disfavor these data.  However, it will be important to confirm these planets with additional observations.

\subsection{Attempts to Break the Period Degeneracies}

\subsubsection{Simulated Signals}

We attempted to break the degeneracies between the periods of planets b and c, using two techniques.  First, we considered the method suggested by \citet{dawson10}--which relies on the periodogram peaks and phases of the signals in comparison to simulated time series--to examine the period of planet c.  Our application of this technique involves first computing a pair of 3-planet Keplerian fits, one for each candidate period for planet c.  Then, using the parameters derived for planet c, we generated a simulated time series for a single planet at the candidate period, using the time stamps and error bars of the original HARPS RVs.  At each time $t_i$ in the simulated series, we added a random perturbation drawn from a Gaussian distribution with $\sigma = \sqrt{\sigma_{RV,i}^2 + (1.8~\textrm{m s}^{-1})^2}$, since our 3-planet models yielded a typical ``jitter" value of 1.8 \mpsp.  We computed the periodogram of the simulated series, and compared to the GLS of the original RVs after removing planets b and d.  Finally, we fit a Keplerian to the simulated data using the ``wrong" period (i.e. we fit a 3.06-day Keplerian to the 0.75-day simulated signal) and compared the phase to the modeled phase of the real data.  Here, we define ``phase" as the mean anomaly ($M_0$) at the first HARPS epoch.

\begin{figure}
\centering
\includegraphics[width=\columnwidth]{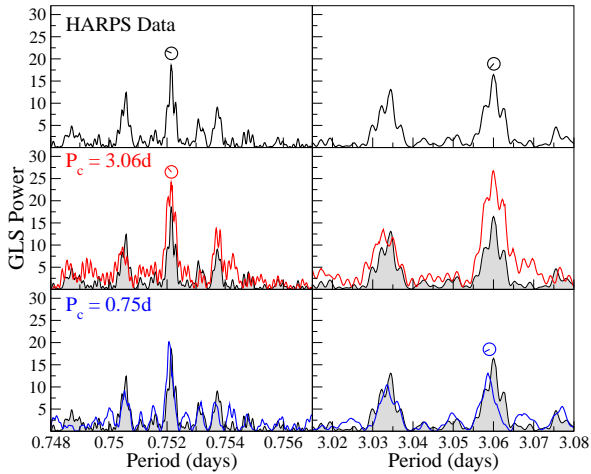}
\caption{\footnotesize The GLS periodogram of the YZ Ceti RVs (\emph{black/gray}) after subtracting planets b and d, compared to simulated time series for planets at 3.06 days (\emph{red, middle row}) and 0.75 days (\emph{blue, bottom row}).  The periodogram of the original data is shown in all three rows for visual comparison.  The dials above each peak show the phase ($M_0$) derived by modeling a Keplerian at that period.  We do not show dials at peaks for which we fixed the planet's phase.}
\label{fig:dawson}
\end{figure}

We performed this exercise for models using periods of 1.97 and 2.02 days for planet b.  In Figure \ref{fig:dawson}, we show the results for the test using $P_b = 2.02$ days, although the choice of $P_b$ makes little difference.  We show the version using $P_b = 2.02$ primarily because it exhibits the greatest discrepancy between the two hypotheses.  For the simulated signals, we find that each candidate period creates significant periodogram power at the 1-day alias, and that the phases of models to the ``wrong" period are similar to those derived from the original RVs.  The one significant difference we observe between the periodograms of the simulated data and the original is that the periodogram power at the true period of the simulated 3-day planet is significantly stronger than that of the original data.  We note, however, that this could be caused by random fluctuation due to the jitter added to the simulated data.  Alternatively, it could indicate incompatibility of the 3.06-day period with the 2.02-day period.  In general, we find that this test does not unambiguously distinguish between the candidate periods, since a planet at either period can create power at both periodogram peaks, and result in Keplerian models that match the phases derived from fitting to incorrect periods.  Furthermore, we elected not to repeat this experiment to distinguish between the candidate periods of planet b because of the greater ambiguity of planet c's period.

\subsubsection{Model Comparison}

Another way to determine which periods are preferred by the current data is to check goodness-of-fit statistics for models to the RVs.  We fit Keplerian models to each of the candidate period combinations using the Markov Chain Monte Carlo (MCMC) RV modeling package \texttt{RadVel} \citep{fulton18}.  For each model, we set the Keplerian orbital elements of each planet, zero-point offsets and additional white-noise terms (``jitters") for the HARPS velocities before and after the fiber upgrade, as free parameters.  The MCMC chains use 150 random walkers, and up to 100\,000 iterations, although the calculation stops if the chains are found to have converged as determined by yielding a value of the Gelman-Rubin statistic less than 1.003 \citep{ford06}.  Where possible, we have retained the model priors used by \citet{astudillo-defru17}.  When evaluating models with $P_c = 0.75$ days, we shifted the period prior to uniform between 0.5 and 1 day.

Prior to computing the MCMC models, we excluded two RVs (BJD = 2457258.798094, 2457606.952332), each of which is more than $6\sigma$ from the data mean.  We removed these values primarily to prevent a biased estimate of the velocity offset between the pre- and post-upgrade RVs.

We evaluated models for solutions of 0, 1, 2, and 3 planets to compare the relative importance of changing the planet periods to that of adding or removing planets.  However, we emphasize that our model comparison is a simple exercise intended primarily to test whether we could distinguish between the aliases for the periods of planets b and c.  A more thorough exploration of the parameter space, including, for example, the impact of various noise models or additional planets, is beyond the scope of this work.

For each model configuration, we compared the BIC and the log of the likelihood ($\ln \La$).  The results of this comparison are summarized in Table \ref{tab:comparison}.  We have assigned values for the false alarm probability (FAP) by comparing the $\ln \La$ values between models with different numbers of planets by using the ``unbiased" likelihood improvement $\tilde{Z}$ from \citet[][Equation 18]{baluev09}.  The FAP then scales approximately as FAP $\propto e^{-\tilde{Z}} \sqrt{\tilde{Z}}$ \citep[e.g.][]{baluev08}.  The FAPs listed in Table \ref{tab:comparison} are relative to the highest-likelihood model with 1 fewer planet than the model under consideration.

Unfortunately, we cannot identify any single model that is clearly preferable to the others.  The FAP values in Table \ref{tab:comparison} show a clear preference for the 2-planet model with $P = 4.66, 0.75$ over the single-planet model, but the case for adding a third planet is marginal compared to the improvement yielded by changing the period of planet c to 18 hours.

Because we used uniform priors in our MCMC models, we may evaluate the relative probabilities of models with the same number of free parameters as $\frac{P_1}{P_2} = e^{\ln \La_1 - \ln \La_2}$.  Thus, for the 3-planet models, we find that the specific configuration proposed by \citet{astudillo-defru17} is approximately 3200 times less likely than our highest-likelihood model, which uses $P_c = 0.75$d.  However, the rest of the models are more similar, with no model above the $\frac{P_1}{P_2} = 150$ threshold typically used as a minimum for unambiguously preferring one model over another \citep[e.g.][]{feroz11}.  Thus, rather than choosing a single ``best" model, we summarize some qualitative results revealed by this analysis:

\begin{itemize}

\item Models with $P_c = 0.75$d are consistently preferred over those with the original 3.06-day period.  The 2-planet solution with $P = 4.66, 0.75$d is clearly the best such model, and both of the highest-likelihood 3-planet models have $P_c = 0.75$d.

\item In light of the 2-planet solution with $P = 4.66, 0.75$d, the addition of a third planet with a period near 2 days is only marginally supported by our analysis.  The FAP for the best 3-planet solution relative to the best 2-planet model is $0.2\%$, suggesting planet b is probably real, but requires additional observations to be confirmed.  Interestingly, the best 2-planet solution was identified in the CSP, suggesting that--as argued by \citet{hara17}--the compressed sensing technique is especially useful for avoiding ambiguities caused by aliasing.

\item Distinguishing between periods of 1.97 and 2.02 days for planet b is particularly difficult.  The 1.97-day period is especially disfavored when adopting the 3.06-day period for planet c or excluding it altogether.  On the other hand, the best model in Table \ref{tab:comparison} uses the 1.97-day period.

\end{itemize}

\begin{table}
\centering
\begin{tabular}{c | c | c c c}
\hline
Number of & Planet Periods & BIC & $\ln \La$ & FAP \\
Planets & (days) & & & \\
\hline
& & & & \\
0 & N/A & 1085.00 & -531.80 & N/A \\
& & & & \\
1 & 4.66 & 1069.90 & -510.87 & $3.7 \times 10^{-6}$ \\
& & & & \\
2 & 4.66, 1.97 & 1082.13 & -503.60 & $>1$ \\
2 & 4.66, 2.02 & 1077.21 & -501.14 & $0.15$ \\
2 & 4.66, 3.06 & 1074.17 & -499.62 & $0.04$ \\
2 & 4.66, 0.75 & 1056.91 & -490.99 & $1.6 \times 10^{-5}$  \\
& & & & \\
3 & 4.66, 1.97, 3.06 & 1069.94 & -484.13 & $>1$ \\
3 & 4.66, 2.02, 3.06 & 1058.94 & -478.63 & $0.02$ \\
3 & 4.66, 2.02, 0.75 & 1057.59 & -477.95 & $0.01$ \\
3 & 4.66, 1.97, 0.75 & 1053.83 & -476.07 & $0.002$ \\
\end{tabular}
\caption{\footnotesize Comparison of goodness-of-fit statistics for our MCMC models to the RV data.  False alarm probability (FAP) is given relative to the highest-likelihood model with 1 fewer planet than the model considered.}
\label{tab:comparison}
\end{table}

\section{Discussion}

In Table \ref{tab:parameters}, we list the modeled and derived parameters for our best fit to the system with planet periods of 4.66, 1.97, and 0.75 days.  The values in Table \ref{tab:parameters} assume a stellar mass $M_* = 0.13 \pm 0.01 M_\odot$, derived from the \citet{delfosse00} $K$-band mass-luminosity relationship \citep[$K = 6.42 \pm 0.02$;][]{cutri03} and the \emph{Gaia} DR2 parallax $\pi = 269.36 \pm 0.08$ mas \citep{gaia_dr2}.  We present this solution not as a replacement for the model presented in \citet{astudillo-defru17}, but rather to serve as a comparison elucidating the consequence of adopting the shorter value of $P_c$.

If the true period of planet c is in fact 0.75 days, it becomes somewhat unique among the known exoplanets.  The minimum masses of the YZ Ceti planets are already the smallest ever discovered with RV, but the revised minimum mass $m_c \sin i = 0.58 M_\oplus$ would establish it as firmly sub-terrestrial in mass.  It would be beneficial to acquire additional RV observations of YZ Ceti during TESS observations in order to better evaluate potential activity contributions to the RVs, and more precisely determine the orbital properties of the low-mass planets in this system.

We used the probabilistic mass-radius prediction routine \texttt{Forecaster} \citep{chen17} to estimate the expected radius of planet c under the assumptions that $P_c = 0.75$ days, and that we are viewing the YZ Ceti system edge-on.  \texttt{Forecaster} predicts a radius $R_c = 0.86 \pm 0.1 R_\oplus$.  The $K$-band radius-luminosity relationship of \citet{mann15} yields a radius $R_* = 0.169 \pm 0.001 R_\odot$ for YZ Ceti, which results in a geometric transit probability of 10\%, more than double the probability derived from the 3.06-day period.  The small stellar radius also yields a relatively high expected transit depth of 0.22\%, which could even potentially be observed from the ground \citep{stefansson17}.  Thus, if the true period of YZ Ceti c is 0.75 days, it offers the potential opportunity for high signal-to-noise study of a transiting sub-terrestrial exoplanet orbiting a relatively bright nearby star.  TESS is currently scheduled to observe YZ Ceti in Sector 3 (September-October 2018) of its survey of the southern hemisphere.  TESS should easily recover the transit signatures of all three planets if they are inclined so as to transit.

If $P_c = 18$ hours, the small orbital separation (high temperature) and small mass (low escape velocity) of the planet could result in significant levels of mass loss.  The planet's atmosphere \citep[e.g. GJ 436b,][]{ehrenreich15} or surface \citep[e.g. KIC 1255b,][]{rappaport12} may be escaping, creating an extended tail of material extending from its surface and causing variable transit depths and durations.  The expected surface temperature of YZ Ceti c at the 18-hour period \citep[$\sim 1000$ K, according to eq. 5 of][]{rappaport12} is too low to vaporize silicates, but tidal forces could result in enhanced volcanic activity that would launch dust from the surface.  Thus, if the planet (or just its exosphere/tail) is transiting, it may provide a unique opportunity to study its atmospheric and interior composition via transit spectroscopy with JWST.

\section{Conclusion}

Our analysis suggests the available HARPS RVs of YZ Ceti are incapable of distinguishing unambiguously between 1-day aliases for the periods of planets b and c.  Our periodograms and model comparisons show a slight preference for revising the period of planet c to 0.75 days, but determining an exact period for planet b is more difficult.  If the period of planet c is in fact 0.75 days, its minimum mass drops to just above half the Earth's mass, and its transit probability increases to 10\%.

\begin{acknowledgements}

This work has made use of observations collected at the European Southern Observatory  under ESO program IDs 180.C-0886(A), 183.C-0437(A), and 191.C-0873(A).  The author is grateful to the anonymous referee for an expeditious and helpful review.  The author also thanks Michael Endl, Gudmundur Stefansson, and Jason Wright for valuable input on this analysis.

\end{acknowledgements}

\begin{table*}
\begin{center}
\begin{tabular}{| l | c c c |}
\hline
Parameter & Planet b & Planet c & Planet d \\
\hline
 & & & \\
 Period $P$ (d) & $1.9689 \pm 4 \times 10^{-4}$ & $0.75215 \pm 1 \times 10^{-5}$ & $4.6568 \pm 4 \times 10^{-4}$ \\
 Time of inferior conjunction $T_C$ (BJD - 245\,0000) & $7662.1 \pm 0.2$ & $7661.56 \pm 0.05$ & $7657.9 \pm 0.2$ \\
 $\sqrt{e} \cos \omega$ & $-0.2 \pm 0.4$ & $-0.2 \pm 0.3$ & $-0.1 \pm 0.3$ \\
 $\sqrt{e} \sin \omega$ & $0.02 \pm 0.3$ & $-0.1 \pm 0.3$ & $0.1 \pm 0.3$ \\
 RV amplitude $K$ (\mpsp) & $1.3 \pm 0.3$ & $1.6 \pm 0.3$ & $1.8 \pm 0.3$ \\
 & & & \\
 HARPS pre-upgrade zero-point offset (\mpsp) & \multicolumn{3}{c |}{$0.1 \pm 0.2$} \\
 HARPS pre-upgrade white-noise jitter $\sigma_{pre}$ (\mpsp) & \multicolumn{3}{c |}{$1.0 \pm 0.3$} \\
 & & & \\
 HARPS post-upgrade zero-point offset (\mpsp) & \multicolumn{3}{c |}{$-0.1 \pm 0.3$} \\
 HARPS post-upgrade white-noise jitter $\sigma_{pre}$ (\mpsp) & \multicolumn{3}{c |}{$1.7 \pm 0.3$} \\
 & & & \\
 \hline
 & & & \\
 Minimum mass $m \sin i$ ($M_\oplus$) & $0.65 \pm 0.15$ & $0.58 \pm 0.11$ & $1.21 \pm 0.2$ \\
 Semi-major axis $a$ (AU) & $0.0156 \pm 4 \times 10^{-4}$ & $0.0082 \pm 2 \times 10^{-4}$ & $0.0276 \pm 7 \times 10^{-4}$ \\
 Eccentricity $e$ & $0.03^{+0.4}_{-0.03}$ & $0.05^{+0.35}_{-0.05}$ & $0.02^{+0.25}_{-0.02}$ \\
 Longitude of periastron $\omega$ ($^\circ$) & $354 \pm 90$ & $30 \pm 80$ & $315 \pm 120$ \\
 \hline
 \end{tabular}
 \caption{\footnotesize Modeled and derived orbital parameters for our best-fit MCMC model to the RVs of YZ Ceti.}
 \label{tab:parameters}
 \end{center}
 \end{table*}

\bibliographystyle{apj}

\bibliography{yzceti_alias.bib}

\clearpage

\end{document}